\documentclass[12pt]{iopart}

\input{packages.sty}
\input{commands.sty}

\begin{document}

\title{Including the vacuum energy in stellarator coil design} 
\author{S.~Guinchard$^{1}\footnote{email: salomon.guinchard@epfl.ch}$, S. R.~Hudson$^{2}$ and E. J.~Paul$^{3}$}
\address{$^1$Swiss Plasma Center, EPFL, Lausanne, VD, 1015, Switzerland}
\address{$^2$Princeton Plasma Physics Laboratory, Princeton, NJ, 08540, USA}
\address{$^3$Department of Applied Physics and Applied Mathematics, Columbia University, New York, NY, 10027, USA}
\vspace{10pt}
\begin{indented}
\item[] Published as \textit{Plasma Phys. Control. Fusion} \textbf{67}, 035028 (2025). \\
\textbf{DOI:} \href{https://doi.org/10.1088/1361-6587/adb789}{10.1088/1361-6587/adb789}
\end{indented}

\begin{abstract}
 Being three-dimensional, stellarators have the advantage that plasma currents are not essential for creating rotational-transform; however, the external current-carrying coils in stellarators can have strong geometrical shaping, which can complicate the construction.
 Reducing the inter-coil electromagnetic forces acting on strongly shaped 3D coils and the stress on the support structure while preserving the favorable properties of the magnetic field is a design challenge.
 In this work, we recognize that the inter-coil forces are the gradient of the vacuum magnetic energy.
 We introduce an objective functional built on the usual quadratic flux on a prescribed target surface together with a weighed penalty on the vacuum energy. 
 The Euler-Lagrange equation for stationary states is derived, and numerical illustrations are computed using a modern stellarator optimization framework. A study of the effect of the energy functional on the inter-coil forces is conducted and the energy is shown to be a promising quantity in producing coils with low forces.
\end{abstract}

\section*{Introduction}

Stellarators offer significant advantages as compared to tokamaks, in that the confining magnetic field is mostly produced by external current-carrying coils, and possibily permanent magnets \cite{helander_permanent_magnet,Qian_MUSE,HAMMOND2024109127}, and so stellarators are less prone to global disruptions. However, because they are three-dimensional, without a continuous symmetry, stellarators must be carefully designed to provide good plasma confinement. 
Similarly, stellarators must be designed to have simple-to-build coils. 
Previous experimental projects such as W7X and NCSX encountered difficulties in building the coils due to their strong shaping\cite{NCSX_report, Risse_W7X}.

 Stellarator optimization traditionally involves two stages \cite{Henneberg_2021}. In the first stage, the shape of the plasma is optimized to obtain favorable confinement; and in the second stage, given the desired plasma boundary, the shape of the external coils is determined.
 More recently, so-called single-stage optimization methods have been introduced \cite{giuliani2022singlestagegradientbasedstellaratorcoil, giuliani_single_stage_II, smiet2024efficientsinglestageoptimizationislands}.
 In this paper, we address the stage-two coil-design problem, for which a desired magnetic surface is provided and the task is to construct a suitable coil set. 
 For simplicity, we consider that the magnetic field is produced entirely by the external set of coils.
 
 Since the pioneering work of Merkel \cite{Merkel_1987}, the coil design problem has been formulated as a minimization problem, where the quantity to be minimized is the integrated squared normal field on the target surface, which is known as the quadratic flux.
 If this quantity is exactly zero, then the target surface will be a flux surface of the magnetic field produced by the coils.
 In Merkel's NESCOIL code, the external currents are represented by a continuous sheet-current density on a prescribed winding surface. The continuous current density can be approximated by a discrete set of filamentary coils that are coincident with a discrete set of contours of the current potential.
 
 Minimizing the quadratic flux alone is not sufficient, for two reasons. The first reason is that minimizing the quadratic flux alone does not lead to a well-defined optimization. Intuitively, for any coil set, another coil set that produces exactly the same magnetic field can be constructed by adding current filaments that are exactly equal and opposite. At the expense of deforming a given coil set in such a way as to mimic a pair of almost equal-and-opposite current filaments, small variations in the magnetic field can be produced that may further reduce the quadratic flux. For this reason, Landreman introduced the REGCOIL code \cite{REGCOIL}, which includes a regularization of the current density.
 
 The second reason is that the coils must satisfy certain engineering constraints; for example, the coils must not be too close to the plasma or to each other, and the inter-coil electromagnetic forces on the coils cannot be too large.
 The stress on the support structure is also a factor to be taken into account when designing a fusion reactor. 
 The stress on the structure can be shown to be linked to the squared magnitude of $\bB$, and thus intrinsically related to the vacuum-field energy \cite{TamuraYanagi}. 
 Cost is also a consideration, and the bigger the structure, the more expensive it is to build\cite{Cost_paper, structure_cost}.
 Moreover, the total mass of the support structure can be expressed, through the virial theorem, as proportional to the stored magnetic energy \cite{takahata2005superconducting, TamuraYanagi} according to 
\beq
M=\frac{Q\rho E}{\sigma_Y},\label{eq:Mass_eq}
\eeq
with $Q$ a correction factor $\sim 3$, $\sigma_Y$ the allowable stress, $\rho$ the density of the structural material and $E$ the stored magnetic energy.

 In this paper, we follow Zhu {\it et al.} \cite{Zhu_2018} and represent the coils as a discrete set of filaments that can move freely in space. We consider how the above-mentioned problems in coil design manifest themselves in the mathematics by considering the possible solutions to the Euler-Lagrange equation for minimizing the quadratic-flux alone. 
 One previously implemented approach \cite{Zhu_2018,HUDSON20182732} for regularizing the coil-optimization problem is to include a penalty on the total length of the coils. In this paper, we introduce an alternative regularization term that is directly related to the inter-coil electromagnetic forces, which are shown to be the  shape-gradient of the magnetic energy.
 We explore to what extent that including the magnetic energy in the coil optimization objective functional serves to both regularize the coil optimization problem and to reduce the inter-coil forces. The method is implemented in SIMSOPT\cite{simsopt, simsopt_energy_release}. 

\section{A minimal variational problem for coil design}

Constructing a minimal, well-posed mathematical problem for stellarator coil design requires careful consideration. 
A necessary requirement is to recover a target magnetic configuration, as defined by a surface, which we shall denote by $S$. 

The standard approach as used in stellarator coil-design codes\cite{ROSE, COILOPT, REGCOIL} is to minimize
\beq
\Phi_2 := \frac{1}{2}\int_{S} dS\;(\bB\cdot \bn)^2,\label{eq:quadratic_flux}
\eeq
which is called the quadratic flux. In Eq.(\ref{eq:quadratic_flux}) $\bB$ denotes the magnetic field and $\bn$ the unit normal to the surface $S$.
For coil design, the degrees of freedom of the system are those that describe the coils, namely the currents that flow through them and their geometry in space. 
In this paper, the filamentary description of the coils as adopted in FOCUS \cite{Zhu_2018} is used, that is they will be described by curves embedded in the 3D space. Only when necessary, such as when evaluating the self-fields and self-forces for examples, some considerations on the finite thickness of the coils will be added. 

We consider the coils to be described by a set of curves $\{C_i\}$ being parameterized by the smooth, periodic vector-valued functions 
\beq
\begin{split}
\bx_i :&[0,\cL_i]\rightarrow \mathbb{R}^3\\
&\ell \mapsto \bx_i(\ell),
\end{split}
\eeq
$\ell$ being the arclength parameter and $\cL_i$ the total length of the coil $i$. 
With the objective being to find a coil configuration that minimizes $\Phi_2$, the geometry of each coil can be changed by a variation $\delta \bx_i$ until the minimal configuration is reached as a solution of a particular Euler-Lagrange (EL) equation. 
To evaluate how the functional $\Phi_2$ changes when the geometry of the coils changes, one needs to evaluate the first variation $\delta \Phi_2[\{\delta \bx_i\}] $. From Eq.(\ref{eq:quadratic_flux}), the first variation of the quadratic flux reads
\beq
\delta \Phi_2 = \int_S (\delta\bB\cdot \bn)(\bB\cdot \bn)\; dS.
\eeq
For filamentary coils, the magnetic field and vector potential can be expressed from the Biot-Savart law,
\beq
\bA(\bx) := \frac{\mu_0}{4\pi} \sum_{i=1}^{N_C}I_i \oint_{C_i}\frac{d\boldsymbol{\ell}}{|\bx-\bx_i|}.\label{eq:Biot-Savart}
\eeq
The first variation is 
\beq
\delta \bA(\bx) 
=
\frac{\mu_0}{4\pi}
\sum_i I_i \oint d\ell \;
\frac{\delta \bx\times(\brr_i\times\bx_i^\prime)}{|\brr_i|^3},
\eeq
where $\brr_i = \bx-\bx_i$. 
In addition, $\delta \bB = \nabla \times \delta \bA$, 
\beq
\delta \bB(\bx) = \frac{\mu_0}{4\pi}\sum_i  I_i\oint d\ell\; (\delta \bx_i\times \bx_i') \cdot \mathcal{R}_i,
\eeq
where $\mathcal{R}_i:=3\brr_i \brr_i/|\brr_i|^5-\mathbb{I}/|\brr_i|^3$. 
The variation of the quadratic flux is then
\beq
\delta \Phi_2 = \frac{\mu_0}{4\pi}\sum_i I_i\oint d\ell \left( \bx_i'\times \int_S\mathcal{R}_{i}^{n}B^n\;dS \right)\cdot\delta \bx_i,\label{eq:deltaPhi2}
\eeq
where $\mathcal{R}_{i}^{n} = \mathcal{R}_i\cdot \bn$ and $\bB\cdot \bn=B^{n}$.  
The coils configuration that minimizes $\Phi_2$ is found from $\delta \Phi_2=0$. 

Eq.(\ref{eq:deltaPhi2}) exhibits a trivial solution, namely $\bB\cdot \bn\equiv0$. Such an ``exact'' solution can be constructed as follows: consider that a coil set is given, and that a flux surface of the corresponding magnetic field is constructed by fieldline tracing or by some other means. If we take that flux surface as the input target surface and initiate the coil optimization procedure, then clearly there is a coil set that exactly produces the target surface. 
Typically however, for a target surface that is produced as part of the plasma optimization procedure (that is, for a target surface that is produced by a ``stage one'') and for which an exact coil set is not provided {\it a priori}, we must consider the case where $B^n$ is not exactly zero everywhere on the target surface.

For the general solution, for which we cannot assume that $B^n\equiv0$, from the definition of $\mathcal{R}_i$, one notes that the solution to $\delta\Phi_2=0$ for arbitrary variations $\delta\bx_i$ is when the distance of the coil to the surface grows to infinity, i.e. $|\bx-\bx_i|\uparrow \infty$.\footnote{Indeed: $\mathcal{R}_{i}^{n}=0 \implies \brr_i/r_i\cdot \bn=1/3$ and $\brr_i/r_i=\bn$, $\Rightarrow\!\Leftarrow$. } 

The minimal problem for coil-design, which consists of minimizing the quadratic flux only, is then ill-posed in that the solution is not unique and that both the exact solution and the grows-to-infinity solution are not achievable from a practical point of view. 
Generally, a term that prevents the coils from growing too long has to be added. 
Moreover, note that in absence of currents, the variation vanishes as well, but no magnetic field is produced so this solution is not physically interesting. To prevent this from happening the currents are kept constant in this work.

The most obvious choice is to include a penalty on the length, as is implemented in FOCUS \cite{Zhu_2018, HUDSON20182732}. Define 
\beq 
\mathcal{F}_{\cL} := \frac{1}{2}\int_{S} dS\;(\bB\cdot \bn)^2 + \omega_{\cL} \sum_{i=1}^{N_C} \oint_{C_i}d\ell = \Phi_2+ \omega_{\mathcal{L}} \mathcal{L},
\eeq
with $\omega_{\cL} \in \mathbb{R_+}$ a weight. Provided that acoil configuration that minimizes $\cF_\cL$ exists, the latter is characterized by 
\beq
\delta\left(\Phi_2+ \omega_{\mathcal{L}} \mathcal{L}\right)[\{\delta\bx_i\}] = 0.
\eeq 
Now that  the objective functional comprises two terms, at the minimum the two shape-gradients are equal and opposite. 
The first variation of the length is related to the curvature of the coils:
\beq
\delta \cL[\{\delta \bx_i\}] = - \sum_{i} \int_{0}^{\cL_i} \kappa_i (\delta \bx_i \cdot \boldsymbol{n}_i) \; d\ell,
\eeq
where $\kappa_i$ denotes the local curvature of the $i$-th coil and $\bn_i$ the normal unit vectors along coil $i$. 
We can now derive an Euler-Lagrange equation for the minimal problem $\Phi_2 + \omega_\cL \mathcal{L}$, namely
\beq
\sum_i \oint d\ell \left(  I_i\bx_i'\times \frac{\mu_0}{4\pi}\int_S\mathcal{R}_{i}^{n}B^n\;dS - \omega_{\cL}\kappa_i \bn_i \right)\cdot\delta \bx_i= 0,
\label{eqn:qfluxlengthEL}
\eeq
with $\omega_{\cL}$ the weight on the length. 
The final EL equations for the quadratic flux and the length read
\beq
\bj_i\times \frac{\mu_0}{4\pi}\int_S\mathcal{R}_{i}^{n}B^n\;dS - \omega_{\cL}\kappa_i \bn_i = \boldsymbol{0} \quad  \forall i\in\{1,...,N_C\},\label{eq:EL_L}
\eeq
implying that the integral term in Eq.(\ref{eq:EL_L}) lies along the $i$-th coil binormal direction, and where $\bj_i$ is the current density in the $i$-th coil. This set of equations is known to have local minima \cite{HUDSON20182732}. 
Although the length has been known to be a good regularizer for the coil design problem involving the quadratic flux, it might not only be the only plausible term. \\

A candidate for the regularization is the vacuum magnetic energy E, determined from computing the squared magnitude of $\boldsymbol{B}$ over all space
\begin{equation}
E := \frac{1}{2\mu_0} \int_{\mathbb{R}^3} B^2 dV.
\end{equation}
The coils being filamentary, the energy can be rewritten as the circulation of $\bA$ along the coils
\beq
E = \sum_i^{N_c} \frac{I_i}{2}\oint_{C_i}\boldsymbol{A}\cdot d\boldsymbol{\ell}. \label{vacuum_energy_flux}
\eeq
The energy thus scales with the length of the coils, making it a candidate for replacing the length objective. Using the Biot-Savart expression Eq.(\ref{eq:Biot-Savart}), the magnetic energy is written as
\beq
E = \frac{\mu_0}{8\pi}\sum_{i,j}I_iI_j\oint_{C_i}\oint_{C_j}\frac{d\boldsymbol{\ell}_i\cdot d\boldsymbol{\ell}_j}{|\bx_i - \bx_j|}.\label{eq:energy_biot_savart}
\eeq 
Note that Eq.(\ref{eq:energy_biot_savart}) involves a summation on $i,j$, such that the terms where $i=j$ are singular. 
The singularity is due to the choice of representation for the coils. The finite thickness of the coils will be considered for numerical evaluation of these terms. 
The energy being expressed in terms of the coils geometries $\bx_i$, a weight on the energy can be added to give another coil objective functional, with a new Euler-Lagrange equation. The variation of the magnetic energy as defined in Eq.(\ref{vacuum_energy_flux}) is 
\beq
\delta E[\{\delta \bx_i\}] = \frac{1}{2}\sum_i \oint_{C_i}d\ell\;(\bj_i\times \bB) \cdot \delta \bx_i.
\eeq
It is interesting to note that the shape gradient of the energy is the Lorentz force acting on the coils. 
In a way similar to Eq.(\ref{eqn:qfluxlengthEL}), the EL equation for the quadratic flux and the magnetic energy reads
\beq
\bj_i\times \left[\frac{\mu_0}{4\pi}\int_S\mathcal{R}_{i}^{n}B^n\;dS - \frac{\omega_E}{2}\bB\right] =\boldsymbol{0} \quad  \forall i\in\{1,...,N_C\},
\eeq
with $\omega_E$ a weight on the magnetic energy. The numerical results presented below suggest that this is a sufficient regularization for coil optimization.
Minimizing the energy along with the quadratic flux prevents the coils from growing to infinite length. 
The shape gradient of $E$ being the $\bj\times \bB$ force, intuition leads to think that the energy plays a role in the forces between the coils, and that penalizing the stored magnetic energy might be helpful in reducing the forces on the structure.

\section{Numerical implementation}

To perform the coil design, the SIMSOPT framework \cite{simsopt} is used. The coils are described mathematically by a set of closed curves in space, whose Cartesian components are expressed in the form of a Fourier series
\beq
X(\phi):=\sum_{n=0}^{N_f} C_n^{X}\cos{(n\phi)} + S_n^{X} \sin(n\phi), \label{eq:coil_cartesian}
\eeq
where $\phi \in [0,2\pi)$ is an angle-like parameter, and $N_f$ the order of the series \cite{Zhu_2018}. Although it is a good approximation of coils far from the plasma surface, the filamentary description of the coils causes issues in the evaluation of the self-fields, self-inductances and self-force due to the singularity term in the Biot-Savart law. Therefore, when evaluating quantities on the coils, some non-zero cross section has to be incorporated in the description to re-establish physical consistency \cite{hurwitz2023efficient,landreman2023efficient}. The optimization of the coils is achieved by minimizing a functional, whose degrees of freedom are the coils' Fourier modes $\{ C_n^{X},S_n^{X} \}_{n,X}$. As introduced in the previous section, the functional in question has to be of the form 
\beq
\mathcal{F} := \frac{1}{2}\int_{S} dS\;(\bB\cdot \bn)^2 + \sum_i\omega_if_i, \label{eq:penalty_func}
\eeq
the quadratic flux term being necessary for the final configuration to be the one of a stellarator. The $\omega_i$ are weights associated to the terms so that each term's effect can be controlled. 

The minimization is achieved by means of L-BFGS-B algorithm implemented in the python library \emph{scipy} \cite{LBFGS, scipy_article}. 
For performance purposes, each objective function $f_i$ calculation is vectorized and the derivatives with respect to the degrees of freedom of the system are implemented in the form of Jacobian vector products (JVP). 
This allows fast computation of the gradients. The main terms considered in this work are
\beq
f_{\mathcal{L}} := \frac{1}{2}\max\left(\sum_{i=1}^{N_C} \cL_i -\cL_0, 0\right)^{2},\label{eq:length_penalty_term}
\eeq
where $N_C$ is the number of coils, $\cL_i$ the length of the $i-$th coil, and $\cL_0$ a length threshold so that the total coils' length start being penalized when it exceeds the latter. 
The energy
\beq
f_{E} = E := \frac{1}{2}\sum_{i\neq j}I_iI_j L_{ij}+\frac{1}{2}\sum_iI_i^2L_i\label{eq:energy_penalty_term}
\eeq
is decomposed as a sum of mutual inductances, simply computed from 
\beq
L_{ij}:=\frac{\mu_0}{4\pi}\int_{0}^{2\pi}d\phi\int_{0}^{2\pi}d\tilde{\phi}\;\frac{\bx_i'(\phi)\cdot \bx_j'(\tilde{\phi})}{|\bx_i(\phi)-\bx_j(\tilde{\phi})|},
\eeq
where $ (0,2\pi]\ni t\mapsto \boldsymbol{x}_i(t)\in \mathbb{R}^3$ is the $i-$th coil parameterization, and self inductances, where the regularization technique involving the cross-section of the coils is taken from Landreman \textit{et al.}\cite{landreman2023efficient}
\beq
L_i:=\frac{\mu_0}{4\pi}\int_0^{2\pi}d\phi\int_0^{2\pi}d\tilde{\phi}\frac{\bx_i'\cdot \tilde{\bx_i}'}{\sqrt{|\bx_i - \tilde{\bx_i}|^2+ \delta a b}},
\eeq
with the regularization term 
\beq
\begin{split}
\delta =& \exp{\left( -\frac{25}{6} + k\right)},\\
k =& \frac{4b}{3a}\tan^{-1}\frac{a}{b} + \frac{4a}{3b}\tan^{-1}\frac{b}{a} + \frac{b^2}{6a^2}\ln{\frac{b}{a}} + \frac{a^2}{6b^2}\ln{\frac{a}{b}}\\
&-\frac{a^4-6a^2b^2+b^4}{6a^2b^2}\ln{\left( \frac{a}{b} + \frac{b}{a}\right)}.
\end{split}
\eeq
The arclength variation on each coil 
\beq
f_{\ell} = \sum_{j=1}^{N_C}\sum_{j=1}^{N_q}\mathrm{Var}(\ell_i^j),
\eeq
where $\ell_i^j$ denotes the arclength variation over the interval between the quadrature points $i$ and $i+1$ of the curve. 
$N_q$ is the number of quadrature points, assumed to be equal for each coil. More specifically, given that the $j$-th coil's curve is parameterized by $t:[0,2\pi)\mapsto \boldsymbol{x}_j(t)$, and that $[0,2\pi)$ is partitioned in $N_q-1$ sub-intervals, $\ell_i^j$ is defined as
\beq
\ell_i^j := \frac{1}{t_{i+1}-t_i}\int_{t_i}^{t_{i+1}} |\boldsymbol{x_j}'(t)|\;dt.
\eeq
Penalizing the variance of the arclength variation on each coil enables to avoid pathological parameterizations due to the non-uniqueness of the parameterization Eq.(\ref{eq:coil_cartesian}). Indeed given a continuously differentiable bijection $\alpha:[0,2\pi)\rightarrow [0,2\pi)$, the Cartesian components $X(\alpha(\phi)):[0,2\pi)\rightarrow \mathbb{R}$ describe the same curve. Numerically enforcing that the parameterization is well behaved can be done by constraining the arclength variation, which can be achieved in different ways. We chose to add a penalty on the variance of the arclength variation between the curves' quadrature points, similar to the approach by Wechsung \textit{et al.}\cite{Wechsung_QS}. Alternatively, as done by Hudson \textit{et al.}\cite{Hudson_Guinchard_Sengupta} a Lagrange multiplier function can also be introduced to constrain $\bx^\prime \cdot \bx^{\prime\prime}$. 

Since ultimately the goal of this study is to assess the effect of minimizing the energy on the inter-coil forces, it is necessary to define metrics for the forces. The usual concern when it comes to forces is the maximum of the Lorentz force on the coil \cite{Robin_2022}. Therefore, we will look at the following metric, defined for a coil $C_i$:
\beq
F_1(C_i):=\max{|\bj_i\times \bB|}.
\eeq
Since it is possible to think of a situation where after optimization the maximum of the Lorentz force has not increased, but the force has increased almost everywhere along the coil, it can be interesting to evaluate the integrated Lorentz force along the coil. We then define the following force metric: 
\beq
F_2(C_i) := \frac{1}{\mathcal{L}_i}\oint_{C_i}d\ell\;  |\bj_i\times \bB|.
\eeq
It is normalized by the length of the coil so both metrics have the units of a force/length.\\

In order to verify that the energy computed from the code is indeed the vacuum field energy, the code has been benchmarked against experimental data from ~W7-X, which has about 620 MJ of energy stored in the vacuum field \cite{W7X_coils}. With a cross sectional area of the coils' conducting part chosen to be $N_w\times 16\times 16$ mm$^2$ to match that of W7-X \cite{W7X_coils_II} where $N_w$ is the number of conductor turns, our calculation gives 619 MJ, which is within a 0.2$\%$ interval from the expected value. We used $N_w^{np}=108$ for non-planar coils and $N_w^p=36$ for planar coils. 

As for the minimization, the derivatives of the energy with respect to the coils' degrees of freedom have been implemented in the form of JVP, with automatic differentiation.

\section{Results}\label{Sec:res}
In the following, the coils have been optimized for the precise Quasi-Axisymmetric (QA) configuration from Landreman-Paul \cite{LandremanPaul2021}. This configuration has two field-periods ($N_{fp}=2$) and is stellarator-symmetric \cite{DEWAR1998275}. We arbitrarily chose $N_C=4$ the number of coils per half field-period. The initial coils were set as circular and equally space along the torus. The currents are initially set in each coil as $0.1$ MA and are kept fixed to prevent the field from vanishing totally and produce a non-physical solution. The same initial configuration has been taken for all the results in this work.

As assumed above, the energy is a valid regularization term, in that a minimizer is found for the functional $\mathcal{F}_E:=\Phi_2 + \omega_E f_E + \omega_{\ell}f_{\ell}$, provided that the weight on the energy is compatible with existence of minimizing solutions. The minimization is achieved with a precision close to machine epsilon, specifically reaching a relative reduction in the objective function  $\leq f\times \text{EPS}$ where $f=1$ and $\text{EPS}$ is the machine epsilon.

\begin{figure}[h!]
\centering
\includegraphics[width = 0.6 \textwidth]{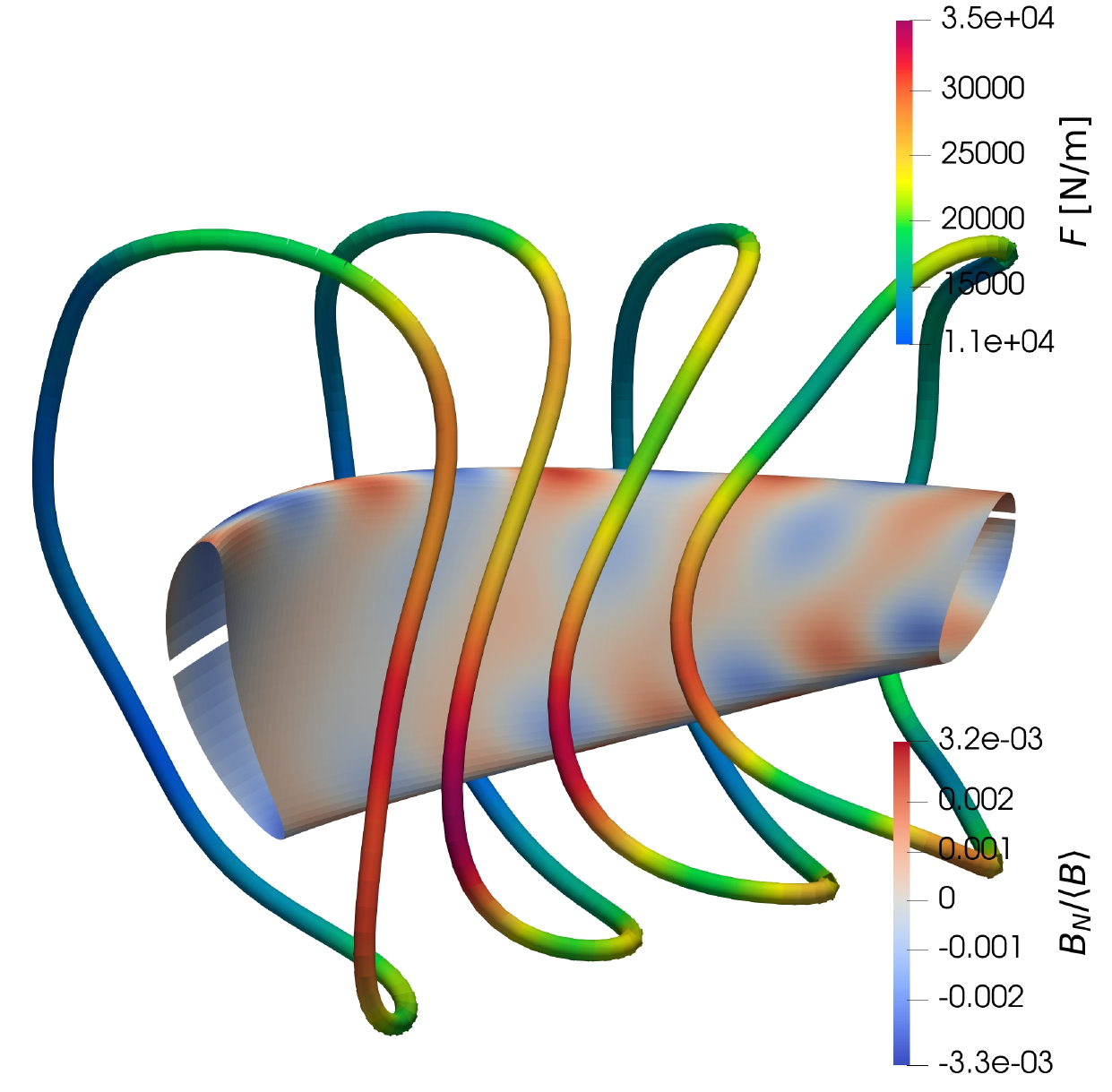}
\caption{Final set of coils produced when the quadratic flux is minimized along with the energy for the Landreman-Paul precise QA target surface. The lower colorbar shows the normalized normal field on the target surface, while the upper colorbar is associated with the forces on the coils. Only a half field period is shown due to stellarator-symmetry.} \label{fig:energy_coils}
\end{figure}

A final configuration for a coil set obtained is shown in Fig.(\ref{fig:energy_coils}). 
The arclength penalty $\omega_{\ell}f_{\ell}$ has been added to get rid of ill-parameterized curves where all the points are pushed back into one region of space, typically the low field side. While conducting numerical computations, it has been observed that as long as $\omega_\ell>0$, the curves' parameterizations remain well-behaved, making $\omega_\ell$ an arbitrary yet fixed parameter. Consequently, when including a penalty on the energy in the objective functional, the weight of the arclength variation penalty does not affect the space of parameters to explore. The characteristics of the final coils are given in Table.(\ref{table:coils_summary}).

The target surface magnetic field is well recovered, as the surface average of the normal field is in the order of $10^{-4}$ Tm$^2$.
The magnetic energy stored in the set of coils is in the order of a few hundreds of kilo-joules. Note that the coils are rather well behaved in that they are separated enough not to exert strong forces on each-other. This behavior may be attributed to the minimization of the mutual-inductance terms in the energy, although some more investigation is necessary to assess wether this is effectively the case. Indeed, the length penalty produces similar coils without explicitly controlling the spacing between the coils. A different magnetic configuration may be necessary for a more thorough analysis. Sufficient spacing of the coils is also required so that the vacuum vessel can be accessed easily and for diverse diagnostics to be inserted. Additionally, the coils achieve a reasonable minimal distance to the surface, which is required for temperature resistance and particle exposure concerns. While it is difficult to make general statements about the energy's ability to control quantities such as coil-coil separation, coil-surface distance, or coil forces — due to the diverse and varied geometries of stellarators and their coils — it seems that in this particular case, the energy minimization tends to regularize these quantities. 

\begin{table}[h!]
\centering
\begin{tabular}{cccccc}
\toprule
& $\cL$ [m] & I [MA] & $\max \kappa$ [m$^{-1}$] & $1/\cL\oint\kappa^2d\ell$ [m$^{-1}$] & $\max |\bj \times \bB|$ [kN/m]\\
$C_1$ & 4.7 & 0.1 & 3.6 & 4.2 & 32.4  \\
$C_2$ & 4.6 & 0.1 & 3.4 & 4.2  & 35.3\\
$C_3$ & 4.4 & 0.1 & 3.8 & 4.6 & 33.3 \\
$C_4$ & 4.4 & 0.1 & 3.9 & 5.2 & 30.5 \\
\midrule 
Global & & $\langle|\bB \cdot \bn| \rangle/\langle B\rangle$ & $E$ [MJ] & $\min d_{CC}$ [m]& $\min d_{CS}$ [m]   \\
quantities& & $9.2 \times 10^{-4}$ & 0.44 & 0.11 & 0.28 \\
\bottomrule
\end{tabular}
\caption{Characteristics of the final set of coils plotted in Fig.(\ref{fig:energy_coils}).}\label{table:coils_summary}
\end{table}

\begin{figure}[h!]
\centering
\includegraphics[width = 1. \textwidth]{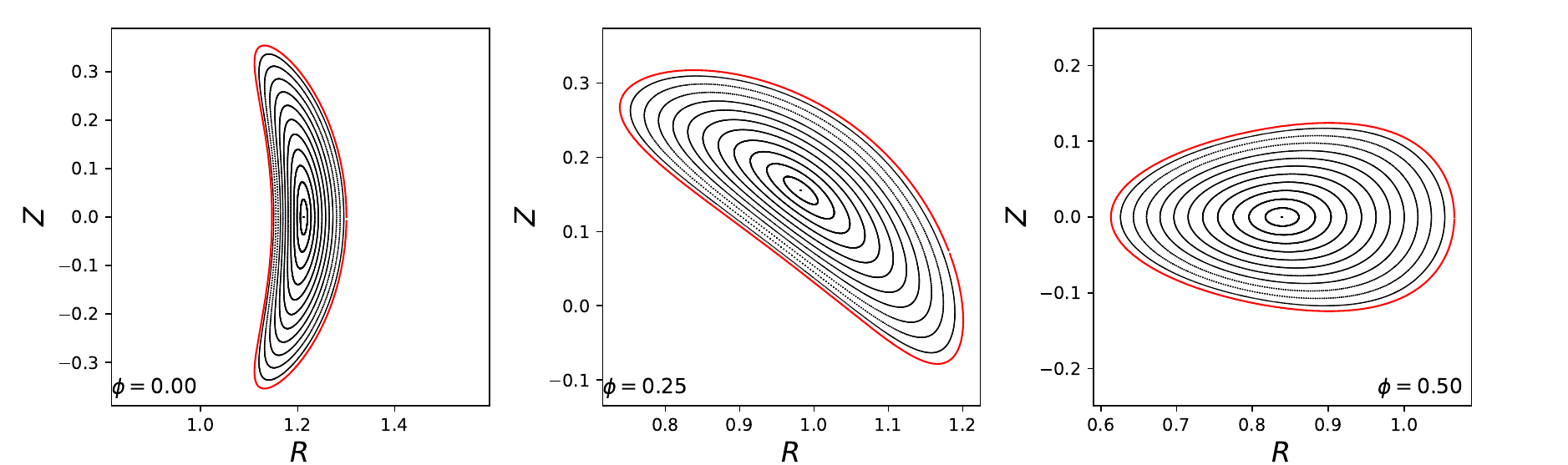}
\caption{Poincaré plot obtained for the set of coils from Fig.(\ref{fig:energy_coils}), at $\phi=0$, $1/4$ and $1/2$ field period. The red curve indicates the boundary targeted by the coils (Landreman-Paul precise QA). The black lines show the field lines traced within the boundary.} \label{fig:Poincare_plot}
\end{figure}

The Poincaré plot for the magnetic field generated by this set of coils is given in Fig.(\ref{fig:Poincare_plot}). The Poincaré sections are shown for 3 different angles within one field-period. This configuration exhibits nicely nested magnetic surfaces within the target boundary. 

The energy and the length are closely related quantities. When penalizing the energy along with the quadratic flux, the length is prevented from diverging. Increasing the weight on the energy in the minimization process will result in coils with shorter length. Such an observation is interesting in that one could eliminate the length term from the objective function, reducing the number of arbitrary weights that need to be assigned. A correlation study between the energy and the total length shows the Pearson coefficient to be $p_{E\cL}=0.99$, giving a very strong linear trend between the two quantities. 
Not only the two quantities exhibit the same trend, but the coils produced from minimizing $\cF_{\cL}=\Phi_2 + \omega_{\cL} f_{\cL}$ and $\cF_{E}=\Phi_2+\omega_{E}f_E + \omega_{\ell} f_{\ell}$ show great geometric similarity, as emphasized in Fig.(\ref{fig:coils_E_L}).

To conduct the correlation study between the quantities of interest, the functional $\cF_E = \Phi_2 + \omega_E f_E + \omega_{\ell}f_{\ell}$ has been minimized for several values of $\omega_E$, keeping $\omega_{\ell}$ constant as its role is only to avoid pathological parameterizations. 
The normalized $\langle \bB\cdot \bn \rangle$, and the two forces metrics have been examined and the correlation coefficients between all quantities are given in Table.(\ref{table:correlation}). As expected, $E$ and $\cL$ exhibit a negative correlation with the normalized normal field. Note that the correlation coefficients between the energy and the two force metrics are positive. 
\begin{table}[h!]
\centering
\begin{tabular}{lccccc}
\toprule
 & $\langle |\bB\cdot \boldsymbol{n}|\rangle/\langle B\rangle$ & $E$ & $\cL$ & $\max_i F_1$ & $\sum_i F_2(C_i)$ \\
\midrule
$\langle |\bB\cdot \boldsymbol{n}|\rangle/\langle B\rangle$ & 1.00 & -0.78 & -0.83 & -0.53 & -0.73 \\
$E$ & -0.78 & 1.00 & 0.99 & 0.78 & 0.98 \\
$\cL$ & -0.83 & 0.99 & 1.00 & 0.76 & 0.97 \\
$\max_i F_1$ & -0.53 & 0.78 & 0.76 & 1.00 & 0.85 \\
$\sum_i F_2(C_i)$ & -0.73 & 0.98 & 0.97 & 0.85 & 1.00 \\
\bottomrule
\end{tabular}
\caption{Pearson's correlation matrix for the normalized averaged normal flux, the energy $E$, the total length $\cL$, and the two force metrics $f_1$ and $f_2$ as $\omega_E$ is scanned.}\label{table:correlation}
\end{table}

One might then question the advantage of using the energy over the length, especially since they appear to yield similar results in this particular case of the precise QA by Landreman and Paul. As shown by Eq. (\ref{eq:Mass_eq}), the total mass of the support structure for the coils is expected to scale with $E$. Notably, for the two coil sets in Fig. (\ref{fig:coils_E_L}), the total stored energy when penalizing the length instead of the energy was approximately $16\%$ higher. Therefore, the configuration obtained from reducing $E$ may need less matter for the support structure.

As one of the concerns for large-scale machines is the forces that the coils would have to withstand during operation, it is of crucial importance to reduce the inter-coil forces as much as possible. 
The fact that the energy shape gradient is the Lorentz force exerted on the coils leads us to think that penalizing the energy is related to the forces acting on the structure. 
To assess to what extent optimizing a set of coils for the Lorentz force is related to optimizing it for the energy, two target functionals are defined. 
Again, we take the minimal
\beq
\cF_E= \Phi_2 + \omega_E f_E + \omega_{\ell} f_{\ell},
\eeq 
and on the other hand, we build a functional that comprises a penalty on the $\bj\times \bB$ force as done by Hurwitz \textit{et al.} \cite{hurwitz2024electromagneticcoiloptimizationreduced}
\beq
\begin{split}
\cF_F =& \Phi_2 + \omega_{\cL}f_{\cL} + \omega_{\ell} f_{\ell} 
+\frac{\omega_F}{2}\sum_\textrm{coils}\oint |\bj\times \bB|^2 d\ell\\
&+\omega_\textrm{cc}\sum_\textrm{coils}\max(d_\textrm{cc,0}-d_\textrm{cc},0) 
+\omega_\textrm{cs}\sum_\textrm{coils}\max(d_\textrm{cs,0}-d_\textrm{cs},0) \\
&+\frac{\omega_\kappa}{2}\sum_\textrm{coils}\oint \text{max}(\kappa - \kappa_0, 0)^2 \textrm{d}\ell
+ \omega_{\kappa_\textrm{MS}}\sum_\textrm{coils}\max\left(\frac{1}{\cL}\oint \kappa^2 d\ell - \kappa_{\text{MS},0},0\right)^2,
\end{split}\label{eq:Hurwitz_force_opt_F}
\eeq
where the two terms on the second line penalize the coil-coil distance and the coil-surface distance, with $d_\textrm{cc,0}$ a minimal threshold above which the final $d_{cc}$ is expected and similarly for the coil-surface distance threshold $d_\textrm{cs,0}$. 
The last two terms penalize the maximum curvature and the mean squared curvature so that the coils are smooth enough. 
Note that no penalty on the energy is present in $\cF_F$.

All these terms are included because of their relevance to stellarator coil design; however, with more targets, the optimization of multi-objective functionals becomes more complicated. Hurwitz et al.\cite{hurwitz2024electromagneticcoiloptimizationreduced} describe the construction of the Pareto front that helps understanding the trade-offs in force optimization, and identifying the relative importance of each weight. The goal of this paper is not to conduct an exhaustive study of the force optimization, and we have not investigated the behavior of the coils under a wide range of weights for the terms in  Eq.(\ref{eq:Hurwitz_force_opt_F}). 

\begin{figure}[h!]
\centering
\includegraphics[width = 0.9 \textwidth]{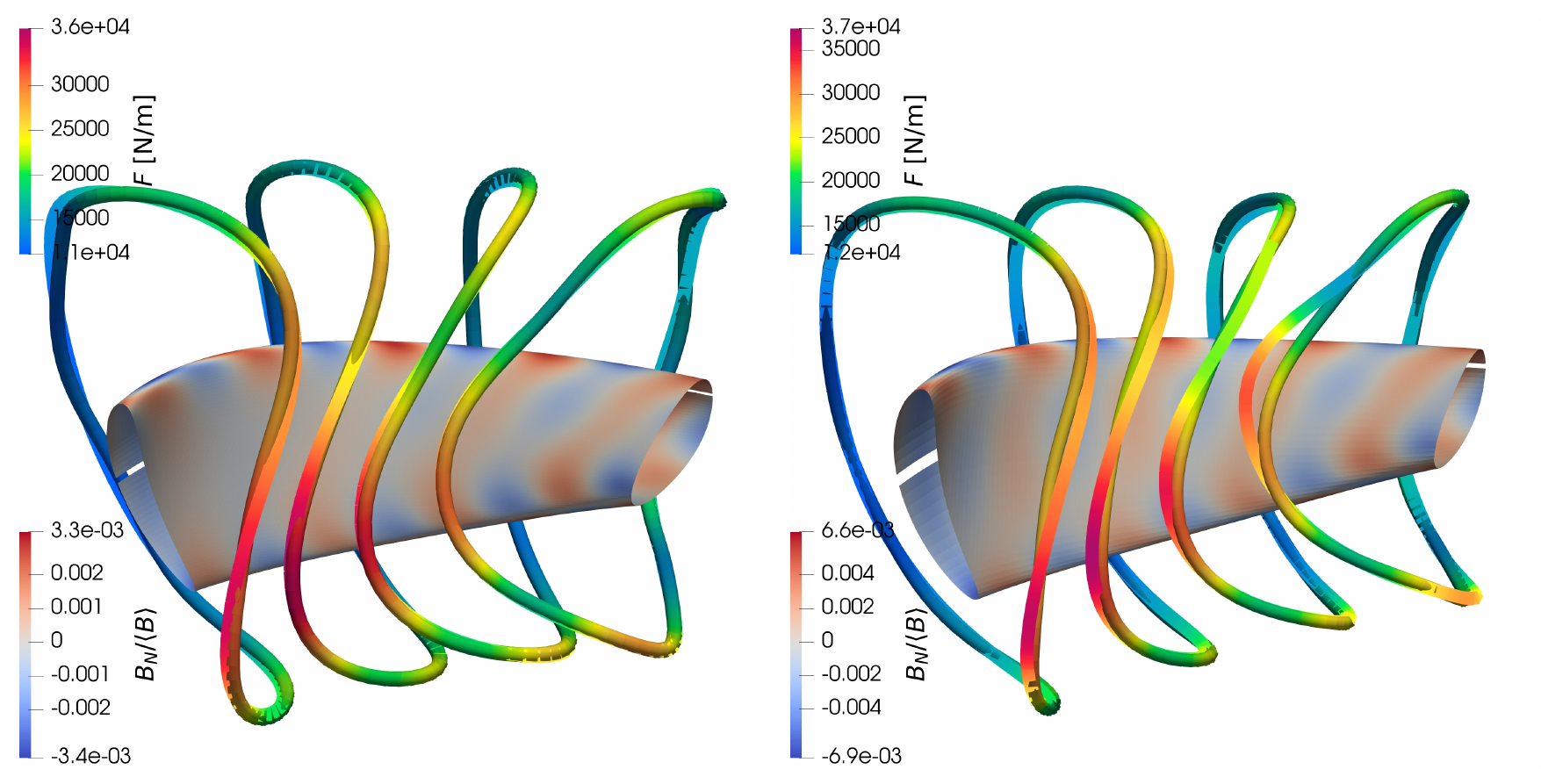}
\caption{Left: coils obtained from minimizing $\cF_E$ with energy weight $\omega_{E,1}$ (matte) superimposed with equivalent set of coils obtained from targeting the functional $\cF_F$ (bright). Right: same for a higher energy weight $\omega_{E,2}$, leading to shorter coils.} \label{fig:coils_E_vs_F}
\end{figure}

The threshold have been set to $d_\textrm{cc,0}=0.1$ m, $d_\textrm{cs,0}=0.3$ m, $\kappa_0 = \kappa_{\text{MS},0}=5$ m$^{-1}$. For two distinct weights on the energy $\omega_{E,i}$, the code is first run with $\cF_E$ and the final total length obtained is then targeted when running the code with $\cF_F$. The weights involved in Eq.(\ref{eq:Hurwitz_force_opt_F}) were chosen arbitrarily according to the threshold considered. The geometric comparisons between the coils obtained from $\cF_E$ and $\cF_F$ are summarized in Table.(\ref{tab:coil_geom}), while the corresponding force results are presented in Table.(\ref{tab:coil_forces}). The coils are plotted in Fig.(\ref{fig:coils_E_vs_F}).

Fig.(\ref{fig:coils_E_vs_F}) shows how similar the final coil configurations are between the two functionals $\cF_E$ (matte curves), and $\cF_F$ (bright curves). 
The coil produced from minimizing $\cF_E$ exhibit a little less regularity on the low field side than those produced with $\cF_F$. This arises because the lower energy penalty constrains less the coil length, allowing for more flexibility in the geometry in areas with low magnetic field. In addition, there are no restrictions on geometric parameters like curvature, which is not penalized in these low-field regions. Moreover, the values of $d_{cc}$, $d_{cs}$ obtained for the coil configurations from $\cF_F$ match those of the given threshold, meaning they had to be penalized and the corresponding terms in Eq.(\ref{eq:Hurwitz_force_opt_F}) were effectively acting. 

As for the forces, the results are very interesting in that for the two chosen weights on the energy, minimizing $\cF_E$ produced coils with lower forces nearly systematically. The results are summarized in the following Table.(\ref{tab:coil_forces}). The bold cells show where the energy approach has reduced the force metrics compared to the $\cF_F$ approach. For the first weight, penalizing the energy reduced the latter by $\sim 10\%$. This enabled to slightly decrease the max force $F_1$ across the four coils. The metric $F_2$ exhibited a slight decrease across three of the four coils. In the $\omega_{E,2}$ case, the energy was reduced by $\sim15\%$, resulting in a reduction of both $F_1$ and $F_2$ on all coils. The max force $F_1$ was reduced by up to $15\%$ and the integrated force $F_2$ saw a reduction of up to $5\%$.

\begin{table}[h]
    \centering
    \begin{tabular}{l c c cccc cccc}
        \toprule
        & E [MJ] & \multicolumn{4}{c}{$F_1$ [kN/m]} & \multicolumn{4}{c}{$F_2$ [kN/m]} \\
        \cmidrule(lr){3-6} \cmidrule(lr){7-10}
        & & Coil 1 & Coil 2 & Coil 3 & Coil 4 & Coil 1 & Coil 2 & Coil 3 & Coil 4 \\
        \midrule
        \multicolumn{10}{c}{\textbf{Case 1:} $\omega_{E,1}$} \\
        \midrule
        \(\mathcal{F}_F\) & $0.49$ & $\bm{34.9}$ &  $\bm{36.5}$ &  $\bm{34.4}$ &  $\bm{30.9}$ &  $\bm{19.6}$ &  $\bm{19.8}$ &  $\bm{19.7}$ & 19.5 \\
        \(\mathcal{F}_E\) & $0.44$ &  $\bm{32.4}$ &  $\bm{35.4}$ &  $\bm{33.3}$ &  $\bm{30.5}$ &  $\bm{19.5}$ &  $\bm{19.6}$ & $\bm{19.6}$ & 19.5 \\
        \midrule
        \multicolumn{10}{c}{\textbf{Case 2:} $\omega_{E,2}>\omega_{E,1}$} \\
        \midrule
        \(\mathcal{F}_F\) & $0.42$ &  $\bm{36.7}$ &  $\bm{37.5}$ &  $\bm{34.9}$ &  $\bm{32.7}$ &  $\bm{20.4}$ &  $\bm{20.7}$ &  $\bm{20.2}$ &  $\bm{20.1}$ \\
        \(\mathcal{F}_E\) & $0.36$ &  $\bm{31.3}$ &  $\bm{32.5}$ &  $\bm{31.9}$ &  $\bm{29.4}$ & $\bm{ 19.4}$ &  $\bm{19.6}$ &  $\bm{19.7}$ &  $\bm{19.7}$ \\
        \bottomrule
    \end{tabular}
    \caption{Force Metrics for \(\mathcal{F}_E\) and \(\mathcal{F}_F\) in the two cases depicted in Fig.(\ref{fig:coils_E_vs_F})}
    \label{tab:coil_forces}
\end{table}

The previously derived results are very promising for producing coils with reduced forces and that require a less voluminous support structure. In the long run, this approach could be cost-saving.

\section{Conclusion}
In this paper, after having described mathematically that the minimal problem for stellarator coil design that consists in minimizing solely the quadratic flux is ill-posed, some regularizing options have been introduced. 
The most obvious and well known that consists in penalizing the length to prevent the coils from growing too large has been briefly reviewed. 
An Euler-Lagrange equation has been derived, linking the change in the quadratic flux to the curvature of the coils. 
This Euler-Lagrange equation was previously derived by Zhu \textit{et al.}\cite{Zhu_2018,HUDSON20182732}.
An alternative approach for regularizing the coil-design problem has been introduced, focusing on the vacuum energy. 
A second Euler-Lagrange equation has been derived, demonstrating that the variation of the quadratic flux has to balance the forces on the coils at the minimizing configuration.

The energy functional has been implemented in the SIMSOPT framework, enabling penalization of the energy at each iteration of the minimization process leading to the coil configuration. 
Results have shown for this configuration to produce regular coils without having to enforce constraints on the length nor on the  coils' curvatures. The coil-coil and coil-surface distances have also been shown to be within acceptable ranges, without having to constrain these terms as well. Penalizing the energy could then have the advantage to provide well-behaved solutions with fewer arbitrary weights in the optimization process. However, while the energy tends to help regularizing those quantities, those control properties should not be considered general. Indeed, results for this magnetic configuration cannot be taken as representative for general stellarators as it is particularly compatible with electromagnetic coils \cite{Kappel_2024}. For stellarators configurations with more shaping or less symmetries, additional constraints on curvature, coil-coil and coil-surface separations are likely to be necessary.   

As for the forces, a correlation study between two forces metrics (the maximum magnitude of the $\bj\times \bB$ force and the integrated force) has shown a positive correlation between the energy and the forces, implying that the forces evolve in the same direction as the energy. 
To assess the efficiency of penalizing the energy at producing low-forces coils, two objective functions were constructed, then minimized, so that the same final length was attained for each set of coils. The final geometries have shown to be very similar, and minimizing the energy has shown a tendency to produce coils with reduced maximal and integrated forces. This latter result is particularly encouraging in that targeting the energy does not work in an opposite direction as optimizing for the forces. In addition, it reduces the structural stress and consequently the required mass for the support structure -- see Eq.(\ref{eq:Mass_eq}).  

To further investigate the influence of the energy on the regularity of the coil configurations and assess more thoroughly the efficiency of penalizing the latter at reducing forces in general, one could decouple the self and mutual inductance components of the energy and weigh them separately. Additionally, level curves of the energy in parameter space could be represented to study the distribution of local extrema. For a more comprehensive study of the effect of energy on forces, one would also need to explore a much wider range of weights for the energy, consider multiple quantities that control and regularize the coils, and trace the Pareto front. Nonetheless, the results presented herein are encouraging.

Furthermore, we hypothesize that the objective functional based on the vacuum energy will have fewer local minima that the objective functional based on the integrated squared force on the coils.
This assumption will be explored in future work.

\section*{Aknowledgments}
The authors would like to thank C. B. Smiet, S. M. Hurwitz and K. Barbey for stimulating discussions and helpful suggestions.  

\section*{Funding}
This work has been carried out within the framework of the EUROfusion Consortium, via the Euratom Research and Training Programme (Grant Agreement No 101052200 - EUROfusion) and funded by the Swiss State Secretariat for Education, Research and Innovation (SERI). Views and opinions expressed are however those of the author(s) only and do not necessarily reflect those of the European Union, the European Commission, or SERI. Neither the European Union nor the European Commission nor SERI can be held responsible for them.\\
This manuscript is based upon work supported by the U.S. Department of Energy, Office of Science, Office of Fusion Energy Sciences, and has been authored by Princeton University under Contract Number DE-AC02-09CH11466 with the U.S. Department of Energy.

\section*{Declaration of interests}
The authors report no conflict of interest.

\section*{Data availability}
The data that support the findings of this study are openly available at the following URL/DOI: \url{https://doi.org/10.5281/zenodo.13935733}\cite{simsopt_energy_release}.

\appendix

\section{Additional results}

\begin{figure}[h!]
\centering
\includegraphics[width = 0.5 \textwidth]{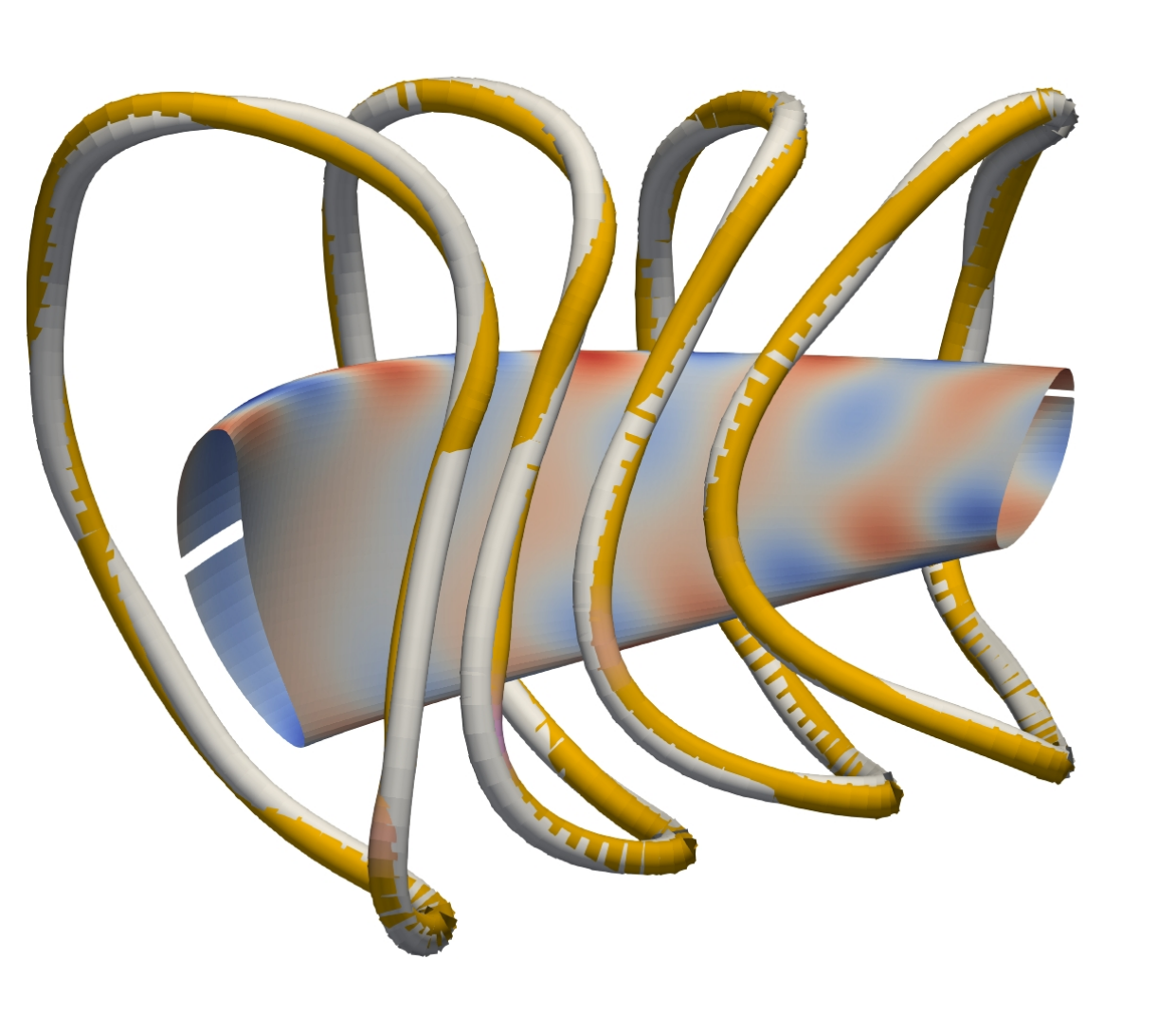}
\caption{White: coils obtained from penalizing the quadratic flux along with the energy. Gold: coils obtained from penalizing the quadratic flux along with the length so that the final length matches the one from the energy optimization.} \label{fig:coils_E_L}
\end{figure}

Fig.(\ref{fig:coils_E_L}) emphasizes the similarity in the geometries of the coils produced from penalizing the length and from penalizing the energy, along with the quadratic flux. Minimizing the energy indeed regularizes the ill-posedness of the coil-design problem from minimizing the quadratic flux.\\

Table.(\ref{tab:coil_geom}) summarizes the geometric properties of the coils obtained from the energy minimization approach and from the forces minimization approach treated in Sec.(\ref{Sec:res}). 
\begin{table}[h]
    \centering
    \begin{tabular}{lccccc}
        \toprule
        & $\mathcal{L}$ [m] & $\max \kappa$ [m$^{-1}$] & $\max \frac{1}{\mathcal{L}} \int d\ell \kappa^2$ [m$^{-1}$] & $\min d_{\text{cc}}$ [m] & $\min d_{\text{cs}}$ [m] \\
        \midrule
        \multicolumn{6}{c}{\textbf{Case 1:} $\omega_{E,1}$} \\
        \midrule
        \(\mathcal{F}_F\) & 18.1 & 3.6 & 4.7 & 0.10 & 0.30 \\
        \(\mathcal{F}_E\) & 18.1 & 3.8 & 5.4 & 0.09 & 0.28 \\
        \midrule
        \multicolumn{6}{c}{\textbf{Case 2:} $\omega_{E,2}>\omega_{E,1}$} \\
        \midrule
        \(\mathcal{F}_F\) & 16.5 & 4.2 & 5.1 & 0.10 & 0.30 \\
        \(\mathcal{F}_E\) & 16.5 & 4.0 & 5.3 & 0.15 & 0.25 \\
        \bottomrule
    \end{tabular}
    \caption{Final geometric parameters for the coils obtained from minimizing $\cF_E$ and $\cF_F$ depicted in Fig.(\ref{fig:coils_E_vs_F}).}
    \label{tab:coil_geom}
\end{table}

\section*{References}
\bibliographystyle{IEEEtran}
\bibliography{references}

\begin{thebibliography}{10}
\providecommand{\url}[1]{#1}
\csname url@samestyle\endcsname
\providecommand{\newblock}{\relax}
\providecommand{\bibinfo}[2]{#2}
\providecommand{\BIBentrySTDinterwordspacing}{\spaceskip=0pt\relax}
\providecommand{\BIBentryALTinterwordstretchfactor}{4}
\providecommand{\BIBentryALTinterwordspacing}{\spaceskip=\fontdimen2\font plus
\BIBentryALTinterwordstretchfactor\fontdimen3\font minus
  \fontdimen4\font\relax}
\providecommand{\BIBforeignlanguage}[2]{{%
\expandafter\ifx\csname l@#1\endcsname\relax
\typeout{** WARNING: IEEEtran.bst: No hyphenation pattern has been}%
\typeout{** loaded for the language `#1'. Using the pattern for}%
\typeout{** the default language instead.}%
\else
\language=\csname l@#1\endcsname
\fi
#2}}
\providecommand{\BIBdecl}{\relax}
\BIBdecl

\bibitem{helander_permanent_magnet}
\BIBentryALTinterwordspacing
P.~Helander, M.~Drevlak, M.~Zarnstorff, and S.~C. Cowley, ``Stellarators with
  permanent magnets,'' \emph{Phys. Rev. Lett.}, vol. 124, p. 095001, Mar 2020.
  [Online]. Available:
  \url{https://link.aps.org/doi/10.1103/PhysRevLett.124.095001}
\BIBentrySTDinterwordspacing

\bibitem{Qian_MUSE}
T.~Qian, X.~Chu, C.~Pagano, D.~Patch, M.~Zarnstorff, B.~Berlinger, D.~Bishop,
  A.~Chambliss, M.~Haque, D.~Seidita, and et~al., ``Design and construction of
  the muse permanent magnet stellarator,'' \emph{Journal of Plasma Physics},
  vol.~89, no.~5, p. 955890502, 2023.

\bibitem{HAMMOND2024109127}
\BIBentryALTinterwordspacing
K.~Hammond and A.~Kaptanoglu, ``Improved stellarator permanent magnet designs
  through combined discrete and continuous optimizations,'' \emph{Computer
  Physics Communications}, vol. 299, p. 109127, 2024. [Online]. Available:
  \url{https://www.sciencedirect.com/science/article/pii/S001046552400050X}
\BIBentrySTDinterwordspacing

\bibitem{NCSX_report}
G.~Neilson, C.~Gruber, J.~Harris, D.~Rej, R.~Simmons, and R.~Strykowsky,
  ``Lessons learned in risk management on ncsx,'' \emph{Plasma Science, IEEE
  Transactions on}, vol.~38, pp. 320 -- 327, 04 2010.

\bibitem{Risse_W7X}
\BIBentryALTinterwordspacing
K.~Riße, ``Experiences from design and production of wendelstein 7-x
  magnets,'' \emph{Fusion Engineering and Design}, vol.~84, no.~7, pp.
  1619--1622, 2009, proceeding of the 25th Symposium on Fusion Technology.
  [Online]. Available:
  \url{https://www.sciencedirect.com/science/article/pii/S0920379609000994}
\BIBentrySTDinterwordspacing

\bibitem{Henneberg_2021}
\BIBentryALTinterwordspacing
S.~A. Henneberg, S.~R. Hudson, D.~Pfefferlé, and P.~Helander, ``Combined
  plasma–coil optimization algorithms,'' \emph{Journal of Plasma Physics},
  vol.~87, no.~2, Apr. 2021. [Online]. Available:
  \url{http://dx.doi.org/10.1017/S0022377821000271}
\BIBentrySTDinterwordspacing

\bibitem{giuliani2022singlestagegradientbasedstellaratorcoil}
\BIBentryALTinterwordspacing
A.~Giuliani, F.~Wechsung, A.~Cerfon, G.~Stadler, and M.~Landreman,
  ``Single-stage gradient-based stellarator coil design: Optimization for
  near-axis quasi-symmetry,'' 2022. [Online]. Available:
  \url{https://arxiv.org/abs/2010.02033}
\BIBentrySTDinterwordspacing

\bibitem{giuliani_single_stage_II}
\BIBentryALTinterwordspacing
A.~Giuliani, F.~Wechsung, A.~Cerfon, M.~Landreman, and G.~Stadler, ``{Direct
  stellarator coil optimization for nested magnetic surfaces with precise
  quasi-symmetry},'' \emph{Physics of Plasmas}, vol.~30, no.~4, p. 042511, 04
  2023. [Online]. Available: \url{https://doi.org/10.1063/5.0129716}
\BIBentrySTDinterwordspacing

\bibitem{smiet2024efficientsinglestageoptimizationislands}
\BIBentryALTinterwordspacing
C.~B. Smiet, J.~Loizu, E.~Balkovic, and A.~Baillod, ``Efficient single-stage
  optimization of islands in finite-$\beta$ stellarator equilibria,'' 2024.
  [Online]. Available: \url{https://arxiv.org/abs/2407.02097}
\BIBentrySTDinterwordspacing

\bibitem{Merkel_1987}
\BIBentryALTinterwordspacing
P.~Merkel, ``Solution of stellarator boundary value problems with external
  currents,'' \emph{Nuclear Fusion}, vol.~27, no.~5, p. 867, may 1987.
  [Online]. Available: \url{https://dx.doi.org/10.1088/0029-5515/27/5/018}
\BIBentrySTDinterwordspacing

\bibitem{REGCOIL}
\BIBentryALTinterwordspacing
M.~Landreman, ``An improved current potential method for fast computation of
  stellarator coil shapes,'' \emph{Nuclear Fusion}, vol.~57, no.~4, p. 046003,
  feb 2017. [Online]. Available:
  \url{https://dx.doi.org/10.1088/1741-4326/aa57d4}
\BIBentrySTDinterwordspacing

\bibitem{TamuraYanagi}
\BIBentryALTinterwordspacing
H.~Tamura, N.~Yanagi, T.~Goto, J.~Miyazawa, T.~Tanaka, A.~Sagara, S.~Ito, and
  H.~Hashizume, ``Mechanical design concept of superconducting magnet system
  for helical fusion reactor,'' \emph{Fusion Science and Technology}, vol.~75,
  no.~5, pp. 384--390, 2019. [Online]. Available:
  \url{https://doi.org/10.1080/15361055.2019.1603041}
\BIBentrySTDinterwordspacing

\bibitem{Cost_paper}
\BIBentryALTinterwordspacing
D.~Chen, J.~Jiang, Y.~Hou, W.~Duan, M.~Ni, and C.~Xing, ``Preliminary cost
  assessment and compare of china fusion engineering test reactor,''
  \emph{Journal of Fusion Energy}, vol.~34, no.~1, pp. 127--132, 2015.
  [Online]. Available: \url{https://doi.org/10.1007/s10894-014-9770-x}
\BIBentrySTDinterwordspacing

\bibitem{structure_cost}
\BIBentryALTinterwordspacing
J.~Sheffield, R.~A. Dory, S.~M. Cohn, J.~G. Delene, L.~F. Parsly, D.~E. Ashby,
  and W.~T. Reiersen, ``Cost assessment of a generic magnetic fusion reactor,''
  \emph{Fusion Technology}, 3 1986. [Online]. Available:
  \url{https://www.osti.gov/biblio/6086979}
\BIBentrySTDinterwordspacing

\bibitem{takahata2005superconducting}
K.~Takahata, ``Superconducting coils for fusion,'' \emph{J. Plasma Fusion Res},
  vol.~81, no.~4, pp. 273--279, 2005.

\bibitem{Zhu_2018}
\BIBentryALTinterwordspacing
C.~Zhu, S.~R. Hudson, Y.~Song, and Y.~Wan, ``New method to design stellarator
  coils without the winding surface,'' \emph{Nuclear Fusion}, vol.~58, no.~1,
  p. 016008, nov 2017. [Online]. Available:
  \url{https://dx.doi.org/10.1088/1741-4326/aa8e0a}
\BIBentrySTDinterwordspacing

\bibitem{HUDSON20182732}
\BIBentryALTinterwordspacing
S.~Hudson, C.~Zhu, D.~Pfefferlé, and L.~Gunderson, ``Differentiating the shape
  of stellarator coils with respect to the plasma boundary,'' \emph{Physics
  Letters A}, vol. 382, no.~38, pp. 2732--2737, 2018. [Online]. Available:
  \url{https://www.sciencedirect.com/science/article/pii/S037596011830759X}
\BIBentrySTDinterwordspacing

\bibitem{simsopt}
\BIBentryALTinterwordspacing
M.~Landreman, B.~Medasani, F.~Wechsung, A.~Giuliani, R.~Jorge, and C.~Zhu,
  ``Simsopt: A flexible framework for stellarator optimization,'' \emph{Journal
  of Open Source Software}, vol.~6, no.~65, p. 3525, 2021. [Online]. Available:
  \url{https://doi.org/10.21105/joss.03525}
\BIBentrySTDinterwordspacing

\bibitem{simsopt_energy_release}
\BIBentryALTinterwordspacing
S.~Guinchard, S.~R. Hudson, and E.~J. Paul, ``Simsopt implementation of vacuum
  energy,'' Oct. 2024. [Online]. Available:
  \url{https://doi.org/10.5281/zenodo.13935733}
\BIBentrySTDinterwordspacing

\bibitem{ROSE}
\BIBentryALTinterwordspacing
M.~Drevlak, C.~Beidler, J.~Geiger, P.~Helander, and Y.~Turkin, ``Optimisation
  of stellarator equilibria with rose,'' \emph{Nuclear Fusion}, vol.~59, no.~1,
  p. 016010, nov 2018. [Online]. Available:
  \url{https://dx.doi.org/10.1088/1741-4326/aaed50}
\BIBentrySTDinterwordspacing

\bibitem{COILOPT}
\BIBentryALTinterwordspacing
L.~A.~B. Dennis J.~Strickler and S.~P. Hirshman, ``Designing coils for compact
  stellarators,'' \emph{Fusion Science and Technology}, vol.~41, no.~2, pp.
  107--115, 2002. [Online]. Available:
  \url{https://doi.org/10.13182/FST02-A206}
\BIBentrySTDinterwordspacing

\bibitem{hurwitz2023efficient}
S.~Hurwitz, M.~Landreman, and T.~M. Antonsen, ``Efficient calculation of the
  self magnetic field, self-force, and self-inductance for electromagnetic
  coils,'' 2023.

\bibitem{landreman2023efficient}
M.~Landreman, S.~Hurwitz, and T.~M. Antonsen, ``Efficient calculation of self
  magnetic field, self-force, and self-inductance for electromagnetic coils.
  ii. rectangular cross-section,'' 2023.

\bibitem{LBFGS}
\BIBentryALTinterwordspacing
R.~H. Byrd, P.~Lu, J.~Nocedal, and C.~Zhu, ``A limited memory algorithm for
  bound constrained optimization,'' \emph{SIAM Journal on Scientific
  Computing}, vol.~16, no.~5, pp. 1190--1208, 1995. [Online]. Available:
  \url{https://doi.org/10.1137/0916069}
\BIBentrySTDinterwordspacing

\bibitem{scipy_article}
\BIBentryALTinterwordspacing
P.~Virtanen, R.~Gommers, T.~E. Oliphant, and et~al, ``Scipy 1.0: fundamental
  algorithms for scientific computing in python,'' \emph{Nature Methods},
  vol.~17, no.~3, pp. 261--272, 2020. [Online]. Available:
  \url{https://doi.org/10.1038/s41592-019-0686-2}
\BIBentrySTDinterwordspacing

\bibitem{Wechsung_QS}
\BIBentryALTinterwordspacing
F.~Wechsung, M.~Landreman, A.~Giuliani, A.~Cerfon, and G.~Stadler, ``Precise
  stellarator quasi-symmetry can be achieved with electromagnetic coils,''
  \emph{Proceedings of the National Academy of Sciences}, vol. 119, no.~13, p.
  e2202084119, 2022. [Online]. Available:
  \url{https://www.pnas.org/doi/abs/10.1073/pnas.2202084119}
\BIBentrySTDinterwordspacing

\bibitem{Hudson_Guinchard_Sengupta}
S.~R. Hudson, S.~Guinchard, and W.~Sengupta, ``Sensitivity of the magnetic axis
  to variations in the magnetic field,'' \emph{Physics of Plasmas}, Oct 2024,
  submitted.

\bibitem{Robin_2022}
\BIBentryALTinterwordspacing
R.~Robin and F.~A. Volpe, ``Minimization of magnetic forces on stellarator
  coils,'' \emph{Nuclear Fusion}, vol.~62, no.~8, p. 086041, jul 2022.
  [Online]. Available: \url{https://dx.doi.org/10.1088/1741-4326/ac7658}
\BIBentrySTDinterwordspacing

\bibitem{W7X_coils}
\BIBentryALTinterwordspacing
M.~Wanner, J.-H. Feist, H.~Renner, J.~Sapper, F.~Schauer, H.~Schneider,
  V.~Erckmann, and H.~Niedermeyer, ``Design and construction of wendelstein
  7-x,'' \emph{Fusion Engineering and Design}, vol. 56-57, pp. 155--162, 2001.
  [Online]. Available:
  \url{https://www.sciencedirect.com/science/article/pii/S0920379601002393}
\BIBentrySTDinterwordspacing

\bibitem{W7X_coils_II}
\BIBentryALTinterwordspacing
K.~Risse, T.~Rummel, L.~Wegener, R.~Holzthüm, N.~Jaksic, F.~Kerl, and
  J.~Sapper, ``Fabrication of the superconducting coils for wendelstein 7-x,''
  \emph{Fusion Engineering and Design}, vol. 66-68, pp. 965--969, 2003, 22nd
  Symposium on Fusion Technology. [Online]. Available:
  \url{https://www.sciencedirect.com/science/article/pii/S0920379603002321}
\BIBentrySTDinterwordspacing

\bibitem{LandremanPaul2021}
\BIBentryALTinterwordspacing
M.~Landreman and E.~Paul, ``Magnetic fields with precise quasisymmetry for
  plasma confinement,'' \emph{Phys. Rev. Lett.}, vol. 128, p. 035001, Jan 2022.
  [Online]. Available:
  \url{https://link.aps.org/doi/10.1103/PhysRevLett.128.035001}
\BIBentrySTDinterwordspacing

\bibitem{DEWAR1998275}
\BIBentryALTinterwordspacing
R.~Dewar and S.~Hudson, ``Stellarator symmetry,'' \emph{Physica D: Nonlinear
  Phenomena}, vol. 112, no.~1, pp. 275--280, 1998, proceedings of the Workshop
  on Time-Reversal Symmetry in Dynamical Systems. [Online]. Available:
  \url{https://www.sciencedirect.com/science/article/pii/S0167278997002169}
\BIBentrySTDinterwordspacing

\bibitem{hurwitz2024electromagneticcoiloptimizationreduced}
\BIBentryALTinterwordspacing
S.~Hurwitz, M.~Landreman, and A.~Kaptanoglu, ``Electromagnetic coil
  optimization for reduced lorentz forces,'' 2024. [Online]. Available:
  \url{https://arxiv.org/abs/2410.09337}
\BIBentrySTDinterwordspacing

\bibitem{Kappel_2024}
\BIBentryALTinterwordspacing
J.~Kappel, M.~Landreman, and D.~Malhotra, ``The magnetic gradient scale length
  explains why certain plasmas require close external magnetic coils,''
  \emph{Plasma Physics and Controlled Fusion}, vol.~66, no.~2, p. 025018, jan
  2024. [Online]. Available: \url{https://dx.doi.org/10.1088/1361-6587/ad1a3e}
\BIBentrySTDinterwordspacing

\end{thebibliography}

\end{document}